\documentclass[aps,prd,superscriptaddress,amsmath,twocolumn,nofootinbib]{revtex4-2}

\usepackage[pdftex]{color}
\usepackage[dvipsnames]{xcolor}
\usepackage[sort&compress]{natbib}
\usepackage{color,hhline,psfrag,rotating,amssymb}
\usepackage[colorlinks=true,linkcolor=PineGreen,filecolor=PineGreen,urlcolor=PineGreen,citecolor=PineGreen,pdftex,plainpages=false]{hyperref}
\usepackage[hang,nooneline]{subfigure}
\usepackage{dcolumn}

\usepackage{graphicx}
\usepackage{revsymb}
\usepackage{dcolumn}

\newcommand{\eq}[1]{Eq.~(\ref{#1})}

\newcommand{\LOHVP}{\mathrm{LOHVP}}
\newcommand{\Ree}{R_{e^+e^-}}
\usepackage{times}

\begin{document}

\title{Utility of a hybrid approach to the hadronic vacuum polarisation contribution to the muon anomalous magnetic moment. }
\author{C.~T.~H.~Davies}\email{christine.davies@glasgow.ac.uk}
\affiliation{SUPA, School of Physics and Astronomy, University of Glasgow, Glasgow, G12 8QQ, UK}
\author{A.~S.~Kronfeld} 
\affiliation{Fermi National Accelerator Laboratory, Batavia, Illinois 60510, USA}
\author{G.~P.~Lepage}\email{g.p.lepage@cornell.edu}
\affiliation{Laboratory for Elementary-Particle Physics, Cornell University, Ithaca, New York 14853, USA}
\author{C.~McNeile}
\affiliation{Centre for Mathematical Sciences, University of Plymouth, PL4 8AA, UK}
\author{R.~S.~Van de Water} 
\affiliation{Fermi National Accelerator Laboratory, Batavia, Illinois 60510, USA}

\date{\today}

\begin{abstract}
 An accurate calculation of the leading-order hadronic vacuum polarisation (LOHVP) contribution to the anomalous magnetic moment of the muon ($a_\mu$) is key to determining whether a discrepancy, suggesting new physics, exists between the Standard Model and experimental results. This calculation can be expressed as an integral over Euclidean time of a current-current correlator $G(t)$, where $G(t)$ can be calculated using lattice QCD or, with dispersion relations, from experimental data for $e^+e^-\to\mbox{hadrons}$. The BMW/DMZ collaboration recently presented a hybrid approach in which $G(t)$ is calculated using lattice QCD for most of the contributing $t$ range, but using experimental data for the largest $t$ (lowest energy) region. Here we study the advantages of varying the position $t=t_1$ separating lattice QCD from data-driven contributions. The total LOHVP contribution should be independent of~$t_1$, providing both a test of the experimental input and the robustness of the hybrid approach.  We use this criterion and a correlated fit to show that Fermilab/HPQCD/MILC lattice QCD results from 2019 strongly favour the CMD-3 cross-section data for $e^+e^-\to\pi^+\pi^-$ over a combination of earlier experimental results for this channel. Further, the resulting total LOHVP contribution obtained is consistent with the result obtained by BMW/DMZ, and supports the scenario in which there is no significant discrepancy between the experimental value for $a_\mu$ and that expected in the Standard Model. We then discuss how improved lattice results in this hybrid approach could provide a more accurate total LOHVP across a wider range of  $t_1$ values with an uncertainty that is smaller than that from either lattice QCD or data-driven approaches on their own. 
\end{abstract}

\maketitle

\section{Introduction}
\label{sec:intro}
As the muon spins its magnetic moment probes the vacuum, interacting with the sea of virtual particles there. These interactions are reflected in the quantity known as the anomalous magnetic moment of the muon, $a_\mu$, recently measured to 0.21 parts per million by the Muon $g-2$ experiment at Fermilab~\cite{Muong-2:2023cdq}. A discrepancy between the experimental value for $a_\mu$ and that calculated in the Standard Model (SM) would indicate the existence of new particles in the virtual sea, beyond those known in the SM. Indeed, such a discrepancy (of size $25(5)\times 10^{-10}$) does exist between the current experimental result and that given in the 2020 Theory White Paper (WP20)~\cite{Aoyama:2020ynm}, with the experimental result being significantly higher. Since 2020, however, it has become clear that the contribution from interactions involving quarks and gluons needs more work. Specifically, the value quoted in WP20 for the largest such contribution, known as the leading-order hadronic vacuum polarisation contribution (LOHVP), has been called into question following several recent developments~\cite{muong-2update}. This makes it important to reassess how the LOHVP is determined, finding ways to maximise the information that is included and embed physical tests of the result where possible. 

The LOHVP can be calculated in two different ways that will be discussed further below: the `data-driven approach' and that using lattice QCD calculations. The LOHVP value quoted in WP20 came from the data-driven approach, using experimental results for the cross-section for $e^+e^-$ annihilation to hadrons as a function of centre-of-mass energy. Using analyticity and the optical theorem, $a_\mu^{\mathrm{LOHVP}}$ is determined as an integral over the cross-section multiplied by a QED kernel function that emphasises low values of $\sqrt{s}$. The WP20 result of $693.1(4.0)\times 10^{-10}$~\cite{Aoyama:2020ynm,Davier:2017zfy,Keshavarzi:2018mgv,Colangelo:2018mtw,Hoferichter:2019mqg,Davier:2019can,Keshavarzi:2019abf} has been assigned a 0.6\% uncertainty that allows for the tension between some of the experimental results available at that time, particularly those seen in the $e^+e^- \rightarrow \pi^+\pi^-$ channel around the $\rho$ resonance which provides the dominant contribution to  $a_\mu^{\mathrm{LOHVP}}$~\cite{akhmetshin:2003zn,aulchenko:2006na,akhmetshin:2006wh,akhmetshin:2006bx,achasov:2006vp,ambrosino:2008aa,ambrosino:2010bv,babusci:2012rp,anastasi:2017eio,lees:2012cj,ablikim:2015orh}. In 2023 a new experimental determination of the $e^+e^- \rightarrow \pi^+\pi^-$ cross-section by the CMD-3 collaboration~\cite{CMD-3:2023alj,CMD-3:2023rfe} gave a result which is larger than previous values. The impact of this CMD-3 result on the combination of experimental results is yet to be fully assessed. Substituting the CMD-3 $2\pi$ cross-section results for the average of other experimental values in the LOHVP, however, closes the gap between the experimental determination of $a_\mu$ and that expected in the SM~\cite{Davier:2023fpl}.

Lattice QCD calculations of the LOHVP can be expressed as an integral over Euclidean time of a correlation function $G(t)$ between two electromagnetic currents. The lattice results for the LOHVP~\cite{FermilabLattice:2017wgj,Budapest-Marseille-Wuppertal:2017okr,RBC:2018dos,Giusti:2019xct,Shintani:2019wai,FermilabLattice:2019ugu,Gerardin:2019rua,Aubin:2019usy,Giusti:2019hkz} at the time of WP20 gave an average of $711.6(18.4)\times 10^{-10}$, with a much larger (2.6\%) uncertainty than the data-driven value. Key problems for the lattice QCD calculations are the degradation of the signal-to-noise ratio in $G(t)$ at large Euclidean times and the determination of finite-volume effects. The large uncertainty meant that the lattice average could not discriminate between the possible scenarios of `no new physics' in $a_\mu$ versus `possible new physics' as favoured by the WP20 data-driven LOHVP. The first subpercent accurate lattice QCD result, by the BMW collaboration~\cite{Borsanyi:2020mff}, was published too late to be included in the WP20 lattice average. The BMW value of $707.5(5.5)\times 10^{-10}$ for the LOHVP has a 0.8\% uncertainty and is $2.1\sigma$ higher than the WP20 data-driven value. 

The BMW collaboration found~\cite{Borsanyi:2020mff} that the significance of the tension between the lattice and data-driven results~\cite{Keshavarzi:2019abf} was increased (to $3.7\sigma$) by considering a partial LOHVP calculation in which the integral over Euclidean time is restricted to a `time-window' (with rounded edges). This had been suggested earlier~\cite{Bernecker:2011gh,RBC:2018dos} as a way to improve the accuracy of lattice QCD calculations by cutting out the noisy large time region.  A more stringent comparison of values from different lattice groups using different discretisations of QCD is then possible and the `intermediate window' between 0.4 fm and 1.0 fm was adopted for this comparison.   There is now a striking agreement on the results for this time-window between multiple different lattice groups with uncertainties at the level of 1--2\%.  Given that results from six different quark formalisms have been used this constitutes one of the best tests of full lattice QCD that has been done.  Results have been obtained for the connected light quark ($u/d$ with $m_u=m_d$) contribution to this window~\cite{Borsanyi:2020mff,Lehner:2020crt,Wang:2022lkq,Aubin:2022hgm,Ce:2022kxy,ExtendedTwistedMass:2022jpw,FermilabLatticeHPQCD:2023jof,RBC:2023pvn,Boccaletti:2024guq}  and for the complete windowed contribution (including all flavours along with quark-line disconnected contributions and QED and $m_u\ne m_d$ corrections)~\cite{Borsanyi:2020mff,Ce:2022kxy,ExtendedTwistedMass:2022jpw,RBC:2023pvn,Boccaletti:2024guq,FermilabLattice:2024yho}.    

The input cross-section data for the data-driven approach can be converted, via a Laplace transform, into a correlation function in Euclidean time~\cite{Bernecker:2011gh} so that a direct comparison with lattice QCD for the time-windowed observables can be made. It is important to note that a disagreement for any time window is as serious as a disagreement for the complete LOHVP contribution and so it makes sense to perform the comparison where the lattice QCD results have comparable accuracy to those from the $e^+e^-$ data. For the 0.4 -- 1.0 fm intermediate time-window, individual lattice QCD results for the complete contribution report tensions between 3.6$\sigma$ and 4.2$\sigma$ with the value from the data-driven average corresponding to the experimental $e^+e^-$ data used in WP20~\cite{Colangelo:2022vok}. The tension is even larger, exceeding 5$\sigma$ for several lattice QCD calculations, if the connected light quark contribution to the window is considered alone~\cite{Benton:2023fcv}. This latter comparison requires analysis of specific hadronic channels in the $e^+e^-$ data to separate out this single flavour contribution~\cite{Benton:2023dci}. Reference~\cite{Benton:2023fcv}  uses the KNT19~\cite{Keshavarzi:2019abf} compilation of experimental $e^+e^-$ data and notes that the replacement in the KNT19 dataset of CMD-3 data for $e^+e^-\to \pi^+\pi^-$~\cite{CMD-3:2023alj} removes the discrepancy between lattice QCD and data-driven results for the connected light quark contribution to the intermediate time-window. This suggests that the lattice QCD results favour the CMD-3 data over earlier experimental measurements of the $e^+e^-\to \pi^+\pi^-$ cross-section.  

Recently the BMW/DMZ collaboration~\cite{Boccaletti:2024guq} have determined a value for the LOHVP contribution using a hybrid technique~\cite{Bernecker:2011gh,RBC:2018dos} in which the Euclidean time axis is divided in two and the result of a lattice QCD calculation from $t=0$ up to 2.8 fm (giving 96\% of the LOHVP value) is combined with a data-driven approach from 2.8 fm to $t=\infty$ (giving the remaining 4\%). This reduces the statistical error from the lattice QCD calculation and the size (and hence uncertainty in) the finite-volume correction, both of which grow at large time. Further, the small size of the data-driven contribution means that even a sizeable relative error in this piece would have little impact. In fact BMW/DMZ stress that they see no tension in the $e^+e^-$ experimental data in the $t>$2.8 fm region, because this is dominated by very low energies below the $\rho$ meson peak. In this way BMW/DMZ aim to reduce the total uncertainty in the LOHVP over that from an equivalent pure lattice QCD result. Their value ($714.1(3.3)\times 10^{-10}$) has an uncertainty of 0.5\%, to be compared with that from earlier lattice QCD calculations discussed above. The LOHVP value obtained by BMW/DMZ leads to the conclusion that the SM and experimental $a_\mu$ results are consistent with each other~\cite{Muong-2:2023cdq,Boccaletti:2024guq}. Further lattice QCD calculations of comparable precision are needed, as well as resolutions to disagreements in the $e^+e^-$ data, to confirm or refute this conclusion. 

To this end we explore more systematically here how to combine lattice QCD and data-driven contributions in an optimal way, modifying and extending a method used in~\cite{RBC:2018dos}. Our method also allows us to test to what extent the input contributions are consistent with each other.  This is equivalent to a comparison of lattice QCD and data-driven results but also leads to a value for the LOHVP contribution. The lattice QCD results that we use to test the approach come from~\cite{FermilabLattice:2022izv}, which used 2019 lattice-QCD data~\cite{FermilabLattice:2019ugu} from the Fermilab Lattice, HPQCD and MILC collaborations, and the compilation of $e^+e^-$ data is based on that of KNT19~\cite{Keshavarzi:2019abf}.

The layout of the paper is as follows: Section~\ref{subsec:hybrid} describes the hybrid approach of combining lattice QCD (subsection~\ref{subsec:lattice}) and $R_{e^+e^-}$ (subsection~\ref{subsec:Ree}) data, before giving the results in subsection~\ref{subsec:results}; Section~\ref{sec:conclusions} gives our conclusions on the utility of this approach and suggestions for future work. Table~\ref{tab:amuR} in subsection~\ref{subsec:Ree} and Tables~\ref{tab:kntcmd} and~\ref{tab:knt} in Appendix~\ref{appendix-corr}  give the values from $\Ree$ needed for future lattice QCD applications of the hybrid approach that could improve uncertainties over the results presented here.

\section{Analysis and Results}
\label{sec:analysis}

\subsection{Hybrid approach}
\label{subsec:hybrid}
The LOHVP contribution to $a_\mu$ can be calculated from the (Euclidean) correlator 
\begin{align}
    G(t) \equiv \sum_{f,f^\prime} Q_f Q_{f^\prime} \sum_{\vec{x}}
    Z_V^2  \langle j_f^i(\vec{x},t) j_{f^\prime}^i(0) \rangle
\end{align}
where $f$, $f^\prime$ label quark flavors, $Q_f$ is the quark's electric charge in units of the electron charge, and $j^i\equiv \overline q \gamma^i q$ is a spatial component of the quark vector current, 
with renormalization factor $Z_V$. Given this correlator, the 
LOHVP contribution is obtained from an integral of the form
\begin{align}
    a_\mu^\LOHVP = \Big(\frac{\alpha}{\pi}\Big)^2
    \int\limits_0^\infty dt \, G(t) K_G(t)
\end{align}
where the kernel $K_G(t)$ vanishes as $t\to0$; see reference~\cite{Bernecker:2011gh} for details.

The correlator $G(t)$ is readily calculated directly using lattice QCD for $t$ values less than 2--3\,fm. Beyond that point, however, statistical errors tend to overwhelm the Monte Carlo in current analyses. The correlator can also be calculated from experimental data for $e^+e^-$  annihilation into hadrons using\,\cite{Bernecker:2011gh}
\begin{align} 
    G_R(t) = \frac{1}{12\pi^2} \int\limits_0^\infty dE\,E^2\,\Ree(E)\,\mathrm{e}^{-Et},
    \label{eq:GR}
\end{align}
where $E=\sqrt{s}$ is the center-of-momentum energy and $\Ree$ is the hadronic cross section divided by the leading order cross section for $e^+e^-\to\mu^+\mu^-$. This suggests a hybrid approach where $G(t)$ is calculated using lattice QCD for small~$t$ values and $\Ree$ data for large~$t$s~\cite{RBC:2018dos}. Of course, this approach assumes there is no new physics beyond the SM in $\Ree$ (see Section~\ref{sec:conclusions}).

To implement this strategy, we divide $a_\mu^\LOHVP$ into two parts,
\begin{align}
    a_\mu^\LOHVP = a_\mu^\LOHVP(0,t_1) + a_\mu^\LOHVP(t_1,\infty),
\end{align}
using a window function  
\begin{align}
    \Theta_{\Delta t}(t_1-t) &= 1 - \Theta_{\Delta t}(t-t_1) 
    \\
    &\equiv 
    \frac{1}{2}\Biggl[1 - \mathrm{tanh}\Big(\frac{t-t_1}{\Delta t}\Big)\Biggr],
\end{align}
which is a step function with a rounded edge of width $\Delta t$:
\begin{align}
    \Theta_{\Delta t}(t_1-t)
    \to
    \begin{cases}
        1 & \mbox{for $t_1-t\gg\Delta t$} \\
        0 & \mbox{for $t-t_1\gg\Delta t$}.
    \end{cases}
\end{align}
Here we set $\Delta t=0.15$\,fm, as in~\cite{FermilabLattice:2022izv}.
We calculate 
\begin{align}
    a_\mu^\LOHVP(0,t_1) \equiv 
    \Big(\frac{\alpha}{\pi}\Big)^2
    \int\limits_0^\infty dt \, G_\mathrm{lat}(t) K_G(t) \Theta_{\Delta t}(t_1-t)
\end{align}
from lattice QCD data that comes predominantly from $t\le t_1$.
Contributions from $t\ge t_1$ are calculated from $\Ree$ data:
\begin{align}
    a_\mu^\LOHVP(t_1,\infty) \equiv 
    \Big(\frac{\alpha}{\pi}\Big)^2
    \int\limits_0^\infty dt \, G_{R}(t) K_G(t) \Theta_{\Delta t}(t-t_1).
\end{align}
Final results for the total $a_\mu^\LOHVP$ should be independent of~$t_1$ within errors, but the error will grow as $t_1$ increases towards 3\,fm, because of statistical errors from lattice QCD.

\begin{figure}
    \includegraphics[scale=0.9]{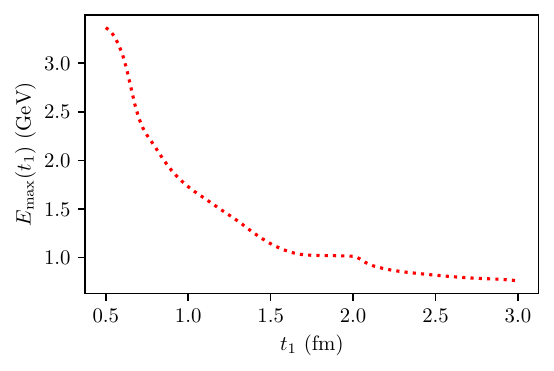}
    \caption{\label{fig:Emax}$E_\mathrm{max}(t_1)$ as a function of $t_1$.
    The bulk of $a_\mu^\LOHVP(t_1,\infty)$ comes from 
    $\Ree$ data with energy $E<E_\mathrm{max}(t_1)$.
    The shoulder around $E_\mathrm{max}=1$\,GeV is 
    due to the strange quark threshold.
    There is a similar shoulder around~3\,GeV from the charm threshold.}
\end{figure}

As $t_1$ increases $a_\mu^\LOHVP(t_1,\infty)$ is dominated by low-energy contributions from $\Ree$. We can demonstrate this by calculating the energy $E_\mathrm{max}(t_1)$
such that the bulk of the contribution to $a_\mu^\LOHVP(t_1,\infty)$ comes from $\Ree$ data with $E\le E_\mathrm{max}(t_1)$. To be precise we define $E_\mathrm{max}$ so that: 
\begin{align}
    a_\mu^\LOHVP(t_1,\infty) - a_\mu^\LOHVP(t_1,\infty)\big|_{E\le E_\mathrm{max}} = 10^{-10},
\end{align}
where $10^{-10}$ is small compared with the current experimental uncertainty in~$a_\mu$ ($2.2\times 10^{-10}$ on the experimental average~\cite{Muong-2:2023cdq}). The value $10^{-10}$ is then a suitable tolerance to allow in the discussion of variations in the corresponding theoretical calculations. 
Figure~\ref{fig:Emax} shows how $E_\mathrm{max}(t_1)$ falls from 3.4\,GeV to 0.76\,GeV as $t_1$ increases from 0.5~to~3.0\,fm, respectively. By varying $t_1$ we probe the energy dependence of~$\Ree$. Lattice QCD and $\Ree$ agree about that energy dependence insofar as results for the total $a_\mu^\LOHVP$ are independent of $t_1$.

\begin{table}
    \caption{\label{tab:amu}
    Lattice QCD results for 
    $a_\mu^\LOHVP(0,t_1)$ from one-sided windows with various $t_1$ values~\cite{FermilabLattice:2022izv}. The
    lattice results are added to results for $a_\mu^\LOHVP(t_1,\infty)$ obtained 
    from the KNT19(CMD-3)~\cite{Benton:2023fcv} and KNT19 datasets~\cite{Keshavarzi:2019abf} for $\Ree$ to obtain 
    estimates for the full contribution $a_\mu^\LOHVP$. We denote the case with no lattice QCD contribution by setting $t_1$ to `None';this is not the same as $t_1=0$ because of the windows rounded edges. } 
    \begin{ruledtabular}
        \begin{tabular}{cccc}
            & $a_\mu^\LOHVP(0,t_1)$ & \multicolumn{2}{c}{$\phantom{ccccc} a_\mu^\LOHVP$} \\
            $t_1$ & LQCD &  + KNT19(CMD-3) &  + KNT19 \\
            \hline
            None & 0 (0)            & 714.7 (4.5)      & 692.7 (2.4)      \\ 
            0.5 & 99.9 (3)         & 715.2 (4.2)      & 694.4 (2.1)      \\ 
            1.0 & 304.0 (1.1)      & 714.0 (3.1)      & 698.8 (1.9)      \\ 
            1.5 & 495.5 (3.5)      & 711.6 (3.9)      & 703.2 (3.7)      \\ 
            2.0 & 605.0 (7.7)      & 705.3 (7.7)      & 701.4 (7.7)     
        \end{tabular}
    \end{ruledtabular}    
\end{table}

\subsection{$a_\mu^\LOHVP(0,t_1)$ from lattice QCD}
\label{subsec:lattice}
Table~\ref{tab:amu} lists results for $a_\mu^\LOHVP(0,t_1)$ taken from 
reference~\cite{FermilabLattice:2022izv}. These are based on lattice QCD calculations,
from 2019, by the Fermilab/HPQCD/MILC collaboration~\cite{FermilabLattice:2019ugu}. They use four sets of configurations with lattice spacings ranging from 0.15~to~0.06\,fm, and $n_f=2+1+1$ flavors of sea quarks. Note that the errors grow steadily as $t_1$~increases.

\subsection{$a_\mu^\LOHVP(t_1,\infty)$ from $\Ree$}
\label{subsec:Ree}
Table~\ref{tab:amuR} lists results for $a_\mu^\LOHVP(t_1,\infty)$ for a range of $t_1$~values. Results are given for two different datasets for $\Ree$. One uses 
the KNT19 dataset\,\cite{Keshavarzi:2019abf}. The other, KNT19(CMD-3), uses the same 
dataset but with the older $e^+e^-\to\pi^+\pi^-$ data replaced with data from CMD-3\,\cite{CMD-3:2023alj}. This latter dataset was generated for Ref.~\cite{Benton:2023fcv}. More details of how this was done from the CMD-3 results are given in the conclusions section of that paper, where the authors also emphasise that the reason for separating the CMD-3 data from earlier data is that the discrepancy seen between them makes a combination currently impossible.

To obtain values for $a_\mu^\LOHVP(t_1,\infty)$, we used \eq{eq:GR} to convert the $\Ree$ data into correlators $G_R(t)$ on a lattice with inverse lattice spacing $a^{-1}=32$\,GeV, which corresponds to a lattice spacing a tenth the size of the smallest lattice spacing used in our simulations (so finite-$a$ errors are negligible). Note that the uncertainties listed in Table~\ref{tab:amuR} are highly correlated; 
see Appendix~\ref{appendix-corr}.

\begin{table}
    \caption{\label{tab:amuR}$a_\mu^\mathrm{LOHVP}(t_1,\infty)$ from KNT19(CMD-3) and KNT19 datasets for $R_{e^+e^-}$ for a range of $t_1$ values. Their difference is given in the final column. The fraction of the 
    total $a_\mu^\LOHVP$ that comes from $a_\mu^\LOHVP(0,t_1)$ (i.e., from lattice QCD)
    is also listed for each $t_1$ value in the second column.}
    \begin{ruledtabular}
        \begin{tabular}{rrrrr}
            $t_1$ &  LQCD Frac. &  KNT19(CMD-3) & KNT19 & Diff. \\
            \hline
            None  &      0.000 &      714.7 (4.5) &      692.7 (2.4) &       21.9 (4.7) \\ 
            0.4   &      0.097 &      645.6 (4.3) &      624.3 (2.2) &       21.3 (4.5) \\ 
            0.5   &      0.139 &      615.3 (4.2) &      594.5 (2.1) &       20.8 (4.4) \\ 
            1.0   &      0.426 &      410.0 (2.9) &      394.8 (1.5) &       15.2 (3.2) \\ 
            1.5   &      0.698 &      216.1 (1.6) &        207.7 (9) &        8.3 (1.8) \\ 
            1.9   &      0.835 &        117.6 (9) &        113.0 (6) &        4.6 (1.1) \\ 
            2.0   &      0.860 &        100.3 (8) &         96.4 (5) &        3.9 (1.0) \\ 
            2.5   &      0.937 &         44.9 (4) &         43.2 (3) &          1.7 (5) \\ 
            2.8   &      0.961 &         27.8 (3) &         26.8 (2) &          1.0 (3) \\ 
            3.0   &      0.972 &         20.4 (2) &         19.6 (2) &          0.7 (3) \\ 
        \end{tabular}
    \end{ruledtabular}
\end{table}
  
\begin{figure}
    \includegraphics[scale=0.9]{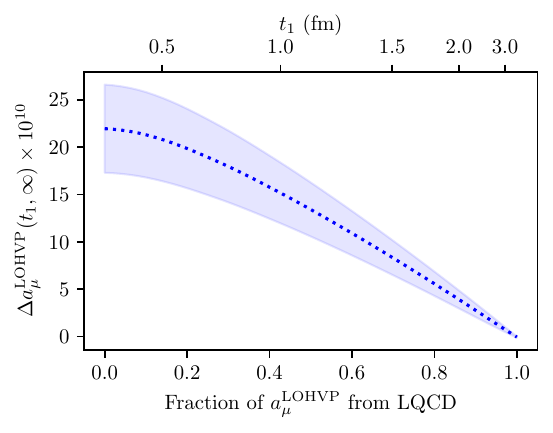}
    \caption{\label{fig:adiff}
    Difference between estimates for $a_\mu^\LOHVP(t_1,\infty)$ made 
    with the KNT19(CMD-3) and KNT19 data sets for $R_{e^+e^-}$. See 
    Table~\ref{tab:amuR} for details.}
\end{figure}

\begin{figure}
    \includegraphics[scale=0.9]{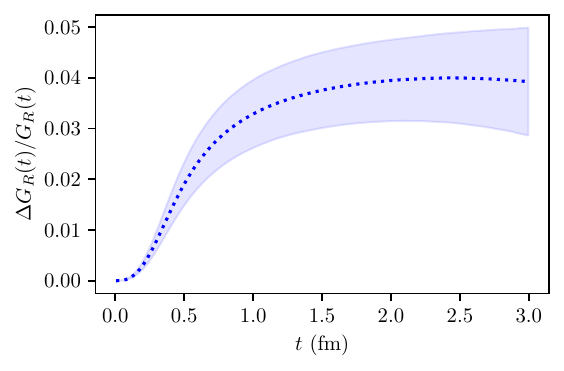}
    \caption{\label{fig:Gdiff}
    Difference $\Delta G_R(t)$ between correlators $G_R(t)$ calculated with 
    the KNT19(CMD-3) and KNT19 datasets, divided by $G_R(t)$ calculated with 
    the KNT19(CMD-3) dataset.}
\end{figure}

The results for $a_\mu^\mathrm{LOVP}(t_1,\infty)$ from KNT19(CMD-3) and KNT19 datasets 
disagree by almost $5\sigma$ at small~$t_1$. Figure~\ref{fig:adiff} shows how this difference decreases with increasing $t_1$ until it is smaller than the experimental uncertainty in $a_\mu$, for $t_1$ of order 2.5\,fm and larger. Note that the difference is still significant ($3\sigma$) at $t_1 =$ 2.8 fm (disagreeing with BMW/DMZ~\cite{Boccaletti:2024guq}), but is of small absolute size ($1\times 10^{-10}$) at that point because the contribution from $R_{e^+e^-}$ is so small. 
Figure~\ref{fig:Gdiff} shows how the differences affect the Euclidean correlator $G_R(t)$, where the discrepancy is of order 4\% for $t$ between 2~and 3\,fm. $\Delta G_R(t)$ falls to zero as $t \rightarrow 0$ because higher energy data (common to both datasets) then dominate Eq.~\eqref{eq:GR}. 

\begin{figure}
    \includegraphics[scale=0.9]{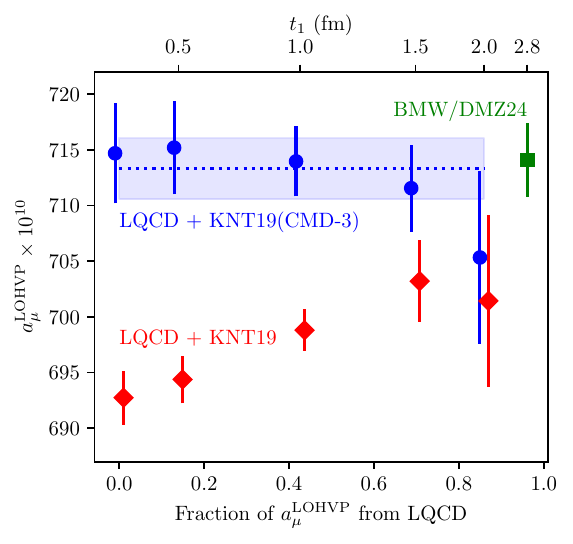}
    \caption{\label{fig:cmd3}Results for $a_\mu^\LOHVP$ from the 
    hybrid approach plotted versus $t_1$. Results are given 
    for two analyses, one that uses the KNT19(CMD-3) dataset for $\Ree$
    (blue circles), and the other that uses the older KNT19 dataset (red diamonds).
    The recent result from BMW/DMZ~\cite{Boccaletti:2024guq} is also shown (green square).
     }
\end{figure}

\begin{table}
    \caption{\label{tab:errorbudget}Error budget showing \%~uncertainties 
    in $a_\mu^\LOHVP$ calculated using the hybrid approach and the 
    KNT19(CMD-3) dataset for $\Ree$. The uncertainties are given for 
    different values of $t_1$ and for the weighted average of the results from 
    different~$t_1$s. The uncertainties come from: the KNT19 dataset for 
    $\Ree$ with contributions from $e^+e^-\to\pi^+\pi^-$ removed; the contribution
    to $\Ree$ from CMD-3's results for  $e^+e^-\to\pi^+\pi^-$; and lattice QCD.  }
    \begin{ruledtabular}
        \begin{tabular}{rcccccc}
          $t_1$ (fm):  & None       & 0.5       & 1.0       & 1.5       & 2.0      & weighted avg. \\ 
    \hline 
        $\Ree$(no $\pi\pi$):  &   0.21   &   0.15  &    0.08  &    0.04   &   0.01  &    0.08 \\
    $\Ree$(CMD-3 $\pi\pi$):  &   0.59   &   0.56  &    0.40  &    0.23   &   0.11  &    0.26 \\
            Lattice QCD:  &   0.00   &   0.04  &    0.15  &    0.50   &   1.09  &    0.27 \\
    \hline
            Total:  &   0.63   &   0.58  &    0.44  &    0.55   &   1.10  &  0.38 \\    
        \end{tabular}
    \end{ruledtabular}
\end{table}

\subsection{Results}
\label{subsec:results}
In Table~\ref{tab:amu} we combine results 
for $a_\mu^\LOHVP(0,t_1)$ from lattice QCD with results for 
$a_\mu^\LOHVP(t_1,\infty)$ from~$\Ree$ to obtain estimates for the total
$a_\mu^\LOHVP$. We do this using each of the KNT19(CMD-3) 
and KNT19 datasets for~$\Ree$.
Again the results from the different datasets do not agree, 
as is clear when they are plotted as a function of~$t_1$ in Figure~\ref{fig:cmd3}.

While the results must agree when extrapolated to large $t_1$, only the results 
from the newer KNT19(CMD-3) are independent of $t_1$ to within errors.
We test the $t_1$ independence of the KNT19(CMD-3)  results (blue circles) by fitting 
them to a constant, taking into account the strong correlations between the points at different $t_1$ values. 
This gives the weighted average of the 
five estimates of $a_\mu^\LOHVP$:
\begin{equation}
    a_\mu^\LOHVP\big[\mbox{LQCD + KNT19(CMD-3)}\big] = 713.3(2.7) \times 10^{-10}
\end{equation}
with $\chi^2/\mathrm{dof} = 0.68$ for $\mathrm{dof}=4$ degrees of freedom ($p$ value 0.6). 
This is an excellent fit, 
unlike what happens when fitting the KTNT19 results (red diamonds):
\begin{equation}
    a_\mu^\LOHVP\big[\mbox{LQCD + KNT19}\big] = 698.6(1.8) \times 10^{-10}
\end{equation}
with $\chi^2/\mathrm{dof} = 3.8$ for $\mathrm{dof}=4$ degrees of freedom ($p$ value 0.005). It is essential in these fits to account for the correlations.

\begin{figure}
    \includegraphics[scale=0.9]{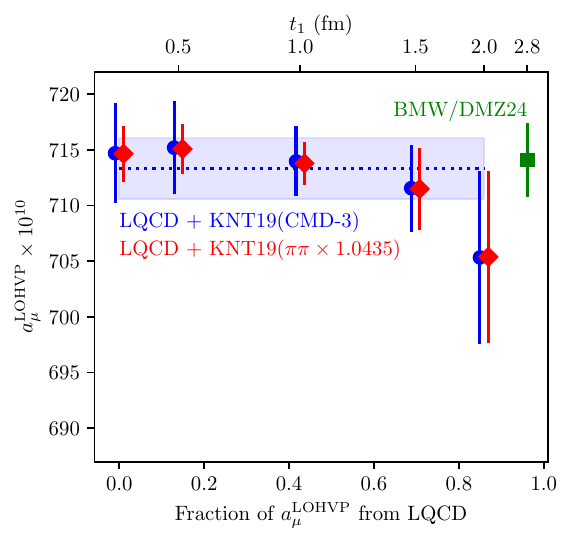}
    \includegraphics[scale=0.9]{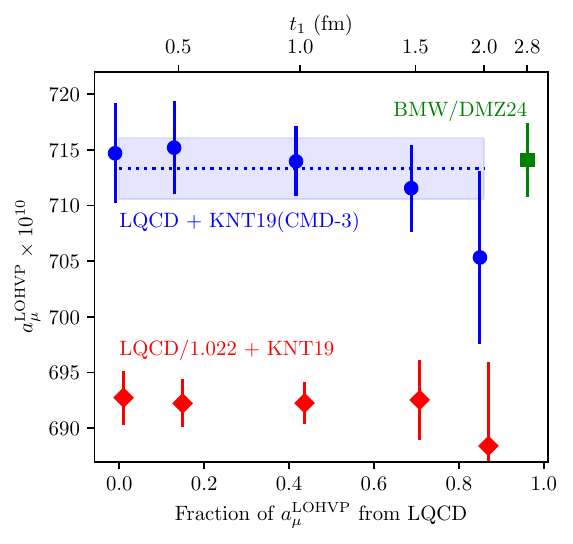}
    \caption{\label{fig:cmd3alt} 
    Results for $a_\mu^\LOHVP$ from the 
    lattice data combined with the KNT19 dataset (red diamonds) 
    versus $t_1$ but with 
    modifications to the $\Ree$ data (top panel) or 
    the lattice results (bottom panel) that make 
    the two consistent with each other.
    In the top panel, KNT19 results for 
    $e^+e^-\to\pi^+\pi^-$ are multiplied by~1.0435.
    In the bottom panel the lattice results are 
    divided by~1.022. Results based on the 
    KNT19(CMD-3) dataset (blue circles) and 
    from BMW/DMZ~\cite{Boccaletti:2024guq}
    (green square) are shown for comparison.
     }
\end{figure}

Focusing on the results from lattice QCD combined with the KNT19 dataset, 
the increase in $a_\mu^\LOHVP$ as $t_1$ increases is roughly proportional to the fraction of $a_\mu^\LOHVP$ coming from lattice QCD. This suggests that the difference between the lattice results and those from the KNT19 data for $\Ree$ is an overall energy-independent multiplicative constant. We can make the KNT19 data consistent with the lattice 
results by multiplying the contributions to $\Ree$ from 
$e^+e^-\to\pi^+\pi^-$ by a factor of~1.0435. This gives a result that is very close 
to that from the KNT19(CMD-3) data (top panel in Figure~\ref{fig:cmd3alt}), but with an uncertainty that is 30\%~smaller. Alternatively we can divide the lattice data by~1.022 (bottom panel of Figure~\ref{fig:cmd3alt}) to obtain a result that is very close to what is obtained from the KNT19 dataset by itself, but with an uncertainty that is 25\% smaller (if we average over all $t_1$). We provide Figure~\ref{fig:cmd3alt} to illustrate the issues arising from the difference between lattice QCD results and KNT19 $\Ree$ data and emphasise that neither of the factors used has any independent justification.

Lattice QCD strongly favors the KNT19(CMD-3) results over those from the KNT19 dataset. The KNT19(CMD-3) results also agree well with the recent result from 
BMW/DMZ~\cite{Boccaletti:2024guq} (green square in Figure~\ref{fig:cmd3}),
and are slightly more accurate when averaged.

Table~\ref{tab:errorbudget} shows the error budget for $a_\mu^\LOHVP$ values 
obtained using the KNT19(CMD-3) dataset. As $t_1$ is increased, the uncertainties 
decrease to 
a minimum around $t_1=1$\,fm. The most accurate estimate comes from the weighted 
average of the results from all $t_1$s, and its uncertainty 
is dominated equally by uncertainties from 
lattice QCD and from the CMD-3 results for $e^+e^-\to\pi^+\pi^-$.

\section{Conclusions}
\label{sec:conclusions}
The choice between a lattice QCD approach and a data-driven approach for calculating $a_\mu^\LOHVP$ 
is not binary. Rather these two approaches are the end-points for a continuum of possibilities labeled 
by $t_1$, the Euclidean time at which we switch from lattice data to $\Ree$ data when calculating 
the current-current correlator $G(t)$ from which $a_\mu^\LOHVP$ is determined. 

There are two reasons for examining the entire $t_1$ dependence, and not just the endpoints or a fixed $t_1$ as in Ref.~\cite{Boccaletti:2024guq}. The 
first is to demonstrate that results are independent of $t_1$ within errors. When this is the case,
it shows that lattice QCD and the $\Ree$ data agree on the energy dependence of $\Ree$. When this 
is not the case, there must be errors in either the lattice QCD analysis or the $\Ree$  data (or both). Our 
analysis shows that the KNT19 dataset for $\Ree$ is not consistent with the lattice QCD data of~\cite{FermilabLattice:2022izv} because results 
are not independent of~$t_1$. The $t_1$ region covered is one where lattice QCD calculations have proven to be reproducible\footnote{See~\cite{FermilabLattice:2022izv} for a comparison with other lattice QCD results in the 0.0--1.0 fm time-window.}. Replacing the old data for $e^+e^-\to\pi^+\pi^-$ in the KNT19 dataset 
with the newer results from CMD-3 yields $t_1$ independence when combined with the lattice QCD data of~\cite{FermilabLattice:2022izv}. These lattice QCD results thereby show a strong preference
for the CMD-3 results over the older data. This agrees with the conclusions of the comparison of intermediate-distance time-windowed results between lattice QCD and $e^+e^-$ data (see Section~\ref{sec:intro} for a discussion) but shows the effect more compellingly with multiple $t_1$ values. 

The $t_1$ independence also strongly suggests that the pure lattice QCD result (in the $t_1 \to \infty$ limit) will agree with the purely data-driven result from KNT19(CMD-3). The larger the range of $t_1$ for which the combination of lattice QCD and $\Ree$ data remains flat the stronger that conclusion becomes as the contribution from the $\Ree$ data becomes smaller. Future lattice QCD calculations will be able to extend the range of $t_1$ values beyond those used here.

A second reason for examining the $t_1$ dependence is that the final uncertainty in $a_\mu^\LOHVP$ 
could well be smaller for intermediate values of $t_1$. This is the case for the KNT19(CMD-3) analysis we present 
here, where the uncertainty is smallest for $t_1=1$\,fm. It is smaller still when we calculate 
the weighted average of the results from different $t_1$s, giving us a value of 
$713.3(2.7) \times 10^{-10}$. This is significantly more accurate than the purely data-driven result for the KNT19(CMD-3) dataset in Table~\ref{tab:amu}. It is also slightly more accurate than the recent BMW/DMZ result of 
$714.1(3.3)\times 10^{-10}$\,\cite{Boccaletti:2024guq}, but agrees well with it and supports the conclusion that there is no significant discrepancy between the experimental results for $a_\mu$ and the Standard Model. 

Much work is needed to come to a definitive answer on $a_\mu$. It is imperative that the discrepancies between the CMD-3 and the older experimental $\Ree$ results 
be understood. 
Considering just the KNT19 data set, the discrepancy between lattice QCD and 
the $\Ree$ data suggests one of three possibilities: 1) there are 3--5\% errors in the KNT19 data 
for $\Ree$ around 1\,GeV or lower; 2) there are 3--5\% errors in the eight lattice results at those energies; 
or 3) there is new physics, beyond the Standard Model, that affects $\Ree$ at the several percent level 
at those energies (this is not easy to arrange - see, for example, Refs.~\cite{DiLuzio:2021uty,Crivellin:2022gfu,Darme:2022yal,Coyle:2023nmi}). 
The CMD-3 data set changes this discussion. If the old data in the KNT19 dataset are correct, then both the 
CMD-3 data set and the lattice QCD results must have errors at the 3--5\% level. And those errors must conspire
to make the hybrid $a_\mu^\LOHVP(t_1)$, calculated from CMD-3 data and lattice QCD results, independent of $t_1$ 
out to at least $t_1=1.5$\,fm. This would be surprising. 

Without understanding the difference between CMD-3 and the older experimental $\Ree$ results one might conclude that a purely lattice QCD result for the LOHVP contribution is the only safe input for the Standard Model value for $a_\mu$ (unless you believe option 2 above, that there are unexplained errors in the lattice QCD results). It is still challenging for multiple lattice QCD groups to achieve $\sim 0.5\%$ uncertainties on the full LOHVP, however (but see recent updates for the full light-quark-connected contribution that achieve 0.7-0.8\% uncertainties in Refs~\cite{RBC:2024fic,Djukanovic:2024cmq,Bazavov:2024eou}). Instead we argue that a combination of lattice QCD and data-driven results is a way forward. Accepting that lattice QCD is correct means that lattice QCD can be used to test the $\Ree$ results, as we have done here, and to select the option that is consistent with $t_1$ independence (if such an option exists). Alternatively it is possible to focus on a large enough value of $t_1$ that the data-driven contribution is small and the systematic error from differences between experiments is not unduly large. This latter approach is the one taken by BMW/DMZ with $t_1 = $ 2.8 fm but, as we show in Table~\ref{tab:amuR}, a smaller $t_1$ value of 2.5 fm or even 2.0 fm can be used without the differences between KNT19 and KNT19(CMD-3) exceeding 1\% of the LOHVP contribution.  

In this way, the uncertainty in the Standard Model prediction could be reduced in future with improved lattice QCD results for a wider range of $t_1$ values. 
In Table~\ref{tab:amuR} we give the numbers for $a_\mu^\LOHVP(t_1,\infty)$ from $\Ree$ to be added to the lattice QCD results for a variety of $t_1$ values to obtain $a_\mu^\LOHVP$ and repeat the analysis done here. The correlation matrices for these numbers are given in Tables~\ref{tab:kntcmd} and~\ref{tab:knt}. The approach suggested here also provides a simple test of new experimental results for $\Ree$. Values for $a_\mu^\LOHVP(t_1,\infty)$ calculated from the new results can be combined with the lattice $a_\mu^\LOHVP(0,t_1)$ values given in Table~\ref{tab:amu} (with correlation matrix in Eq.\eqref{eq:latcorr}), or improved lattice results when they become available. The lattice QCD results used here are from 2019 and more recent lattice data from, for example, Fermilab/HPQCD/MILC, has higher statistics and includes results on finer lattices~\cite{FermilabLatticeHPQCD:2023jof}. From the error budget in Table~\ref{tab:errorbudget} we see that halving the lattice QCD uncertainty at $t_1 =$ 2.0 fm would reduce the uncertainty on $a_\mu^\LOHVP$ to 0.56\% using KNT19(CMD-3) results, or 0.8\% if an additional uncertainty is allowed for the difference between KNT19 and KNT19(CMD-3). This is achievable in the near future.

\begin{appendix}
\section{}\label{appendix-corr}
The uncertainties in the lattice QCD results for $a_\mu^\LOHVP(0,t_1)$ 
in Table~\ref{tab:amu} are correlated. 
The correlation matrix for the results at $t_1=0.5$, 1.0, 1.5, and~2.0\,fm is
\begin{align}
\label{eq:latcorr}
    \begin{pmatrix}
        1.0000 & 0.3110  & 0.1840  & 0.1135 \\
        0.3110 & 1.0000  & 0.5484  & 0.4034 \\
        0.1840 & 0.5484  & 1.0000  & 0.7287 \\
        0.1135 & 0.4034  & 0.7287  & 1.0000 \\               
    \end{pmatrix}
\end{align}
Similarly the uncertainties in $a_\mu^\LOHVP(t_1,\infty)$ listed in Table~\ref{tab:amuR} are (highly) correlated. The correlation
matrices for the two datasets are given in Tables~\ref{tab:kntcmd} and~\ref{tab:knt}. There is no correlation between the lattice QCD and $\Ree$ results. 

\begin{table*}
        \caption{\label{tab:kntcmd}Correlation matrix for KNT19(CMD-3) results in Table~\ref{tab:amuR}.}
    \begin{tabular}{c|cccccccccc}\hline\hline
    $t_1$ & None & 0.4 & 0.5 & 1.0 & 1.5 & 1.9 & 2.0 & 2.5 & 2.8 & 3.0 \\
    \hline
    None & 1.0000 & 0.9949 & 0.9915 & 0.9748 & 0.9588 & 0.9371 & 0.9295 & 0.8758 & 0.8331 & 0.8019 \\ 
    0.4 & 0.9949 & 1.0000 & 0.9995 & 0.9900 & 0.9758 & 0.9544 & 0.9468 & 0.8927 & 0.8496 & 0.8180 \\ 
    0.5 & 0.9915 & 0.9995 & 1.0000 & 0.9935 & 0.9804 & 0.9595 & 0.9520 & 0.8982 & 0.8550 & 0.8234 \\ 
    1.0 & 0.9748 & 0.9900 & 0.9935 & 1.0000 & 0.9947 & 0.9789 & 0.9725 & 0.9241 & 0.8834 & 0.8528 \\ 
    1.5 & 0.9588 & 0.9758 & 0.9804 & 0.9947 & 1.0000 & 0.9942 & 0.9906 & 0.9551 & 0.9209 & 0.8936 \\ 
    1.9 & 0.9371 & 0.9544 & 0.9595 & 0.9789 & 0.9942 & 1.0000 & 0.9995 & 0.9807 & 0.9554 & 0.9328 \\ 
    2.0 & 0.9295 & 0.9468 & 0.9520 & 0.9725 & 0.9906 & 0.9995 & 1.0000 & 0.9861 & 0.9635 & 0.9426 \\ 
    2.5 & 0.8758 & 0.8927 & 0.8982 & 0.9241 & 0.9551 & 0.9807 & 0.9861 & 1.0000 & 0.9942 & 0.9834 \\ 
    2.8 & 0.8331 & 0.8496 & 0.8550 & 0.8834 & 0.9209 & 0.9554 & 0.9635 & 0.9942 & 1.0000 & 0.9971 \\ 
    3.0 & 0.8019 & 0.8180 & 0.8234 & 0.8528 & 0.8936 & 0.9328 & 0.9426 & 0.9834 & 0.9971 & 1.0000 \\    
    \hline
    \hline 
    \end{tabular}
    \end{table*}

    \begin{table*}
        \caption{\label{tab:knt}Correlation matrix for KNT19 results in 
        Table~\ref{tab:amuR}.}
    \begin{tabular}{c|cccccccccc}\hline\hline
    $t_1$ & None & 0.4 & 0.5 & 1.0 & 1.5 & 1.9 & 2.0 & 2.5 & 2.8 & 3.0 \\
    \hline
    None & 1.0000 & 0.9827 & 0.9711 & 0.9053 & 0.8343 & 0.7585 & 0.7376 & 0.6361 & 0.5839 & 0.5536 \\ 
    0.4 & 0.9827 & 1.0000 & 0.9982 & 0.9576 & 0.8907 & 0.8129 & 0.7912 & 0.6846 & 0.6293 & 0.5972 \\ 
    0.5 & 0.9711 & 0.9982 & 1.0000 & 0.9714 & 0.9086 & 0.8317 & 0.8100 & 0.7027 & 0.6469 & 0.6143 \\ 
    1.0 & 0.9053 & 0.9576 & 0.9714 & 1.0000 & 0.9748 & 0.9175 & 0.8993 & 0.8034 & 0.7501 & 0.7180 \\ 
    1.5 & 0.8343 & 0.8907 & 0.9086 & 0.9748 & 1.0000 & 0.9820 & 0.9726 & 0.9093 & 0.8678 & 0.8410 \\ 
    1.9 & 0.7585 & 0.8129 & 0.8317 & 0.9175 & 0.9820 & 1.0000 & 0.9990 & 0.9704 & 0.9430 & 0.9234 \\ 
    2.0 & 0.7376 & 0.7912 & 0.8100 & 0.8993 & 0.9726 & 0.9990 & 1.0000 & 0.9802 & 0.9567 & 0.9391 \\ 
    2.5 & 0.6361 & 0.6846 & 0.7027 & 0.8034 & 0.9093 & 0.9704 & 0.9802 & 1.0000 & 0.9951 & 0.9878 \\ 
    2.8 & 0.5839 & 0.6293 & 0.6469 & 0.7501 & 0.8678 & 0.9430 & 0.9567 & 0.9951 & 1.0000 & 0.9983 \\ 
    3.0 & 0.5536 & 0.5972 & 0.6143 & 0.7180 & 0.8410 & 0.9234 & 0.9391 & 0.9878 & 0.9983 & 1.0000 \\ 
     \hline
    \hline 
    \end{tabular}
    \end{table*}

\end{appendix}  

\subsection*{\bf{Acknowledgements}} 
We are grateful to Alex Keshavarzi for providing the KNT19 and KNT19(CMD-3) datasets. In addition we are grateful to him and to Thomas Teubner for many useful discussions. 
The lattice QCD results used here come from~\cite{FermilabLattice:2019ugu}. We enjoyed stimulating discussions over these results with the co-authors of that work and are grateful for the collaboration. 
The lattice QCD results used the DiRAC Data Analytic system at the University of Cambridge, operated by the University of Cambridge High Performance Computing Service on behalf of the STFC DiRAC HPC Facility (www.dirac.ac.uk). This equipment was funded by BIS National E-infrastructure capital grant (ST/K001590/1), STFC capital grants ST/H008861/1 and ST/H00887X/1, and STFC DiRAC Operations grant ST/K00333X/1. DiRAC is part of the National E-Infrastructure.
We are grateful to the Cambridge HPC support staff for assistance. Computations for the lattice QCD results were also carried out with resources provided by the USQCD Collaboration, the National
Energy Research Scientific Computing Center and the Argonne Leadership Computing Facility, which are funded
by the Office of Science of the U.S.\ Department of Energy.
{The lattice QCD work used the Extreme Science and Engineering Discovery Environment (XSEDE) supercomputer Stampede 2 at the Texas Advanced Computing Center (TACC) through allocation TG-MCA93S002.  The XSEDE program is supported by the National Science Foundation under grant number ACI-1548562.}
{Computations on the Big Red II+ supercomputer were supported in part by Lilly Endowment, Inc., through its support for the Indiana University Pervasive Technology Institute.} The parallel file system employed by Big Red II+ is supported by the National Science Foundation under Grant No.~CNS-0521433.
The lattice QCD results also utilized the RMACC Summit supercomputer, which is supported by the National Science Foundation (awards ACI-1532235 and ACI-1532236), the University of Colorado Boulder, and Colorado State University. The Summit supercomputer is a joint effort of the University of Colorado Boulder and Colorado State University.
Some of the computations were done using the Blue Waters sustained-petascale computer, which was supported by the National Science Foundation (awards OCI-0725070 and ACI-1238993) and the state of Illinois. Blue Waters was a joint effort of the University of Illinois at Urbana-Champaign and its National Center for Supercomputing Applications.

Funding for this work came from the
Science and Technology Facilities Council
and the National Science Foundation.

 \bibliography{hybrid}

\end{document}